\documentclass[12pt]{article}

\usepackage{times}
\usepackage{rotating}
\usepackage{graphicx}
\usepackage{xcolor}

\bibliography{scibib}
\newenvironment{sciabstract}{%
\begin{quote} \bf}
{\end{quote}}

\title{Tunable strain soliton networks confine electrons\\ in Van der Waals materials} 

\author
{Drew Edelberg,$^{1}$ Hemant Kumar,$^{2}$ Vivek Shenoy,$^{3}$ \\ H\'ector Ochoa, $^{1\ast}$ Abhay N. Pasupathy $^{1\ast}$\\
\\
\normalsize{$^{1}$Department of Physics, Columbia University}\\
\normalsize{538 W. 120 Street, New York, NY 10027, USA}\\
\normalsize{$^{2}$School of Basic Sciences, Indian Institute of Technology,}\\
 \normalsize{Bhubaneswar, 752050, India}\\
\normalsize{$^{3}$Department of Materials Science and Engineering, University of Pennsylvania}\\
\normalsize{ 3231 Walnut St., Philadelphia, PA 19104, USA}\\
\\
\normalsize{$^\ast$To whom correspondence should be addressed:} \\
\normalsize{E-mail: ho2273@columbia.edu, apn2108@columbia.edu.}
}

\date{}

\begin{document} 

\baselineskip24pt

\maketitle 

\begin{sciabstract}
Sliding and twisting van der Waals layers with respect to each other gives rise to moir\'e structures with emergent electronic properties. Electrons in these moir\'e structures feel weak  potentials that are typically in the tens of millielectronvolt range when the moir\'e structures are smooth at the atomic scale. Here we report a facile technique to achieve deep, deterministic trapping potentials via strain-based moir\'e engineering in van der Waals bilayers. We use elasto-scanning tunneling microscopy to show that uniaxial strain drives a commensurate-incommensurate lattice transition in a multilayer MoSe$_2$ system. In the incommensurate state, the top monolayer is partially detached from the bulk through the spontaneous formation of topological solitons where stress is relieved. Intersecting solitons form a honeycomb-like network resulting from the three-fold symmetry of the adhesion potential between layers. The vertices of the honeycomb network occur in bound pairs with different interlayer stacking arrangements. One vertex of the pair is found to be an efficient trap for electrons, displaying two states that are deeply confined within the semiconductor gap and have a spatial extent of 2 nm. Honeycomb soliton networks thus provide a unique path to engineer an array of identical deeply confined states with a strain-dependent tunable spatial separation, without the necessity of introducing chemical defects into the host materials.
\end{sciabstract}

Heterostructures of van der Waals materials offer elegant routes for bandstructure engineering simply by placing two van der Waals sheets in contact with each other. When a  mismatch in the lattice vectors of the two sheets is present, a long-wavelength moir\'e determined by the difference between the two lattice vectors is imposed on the system. Such mismatches can exist for several reasons: differences between the intrinsic lattice constants of the two layers as is the case for graphene on BN \cite{hofstader};  rotations between the two lattices as is the case for twisted bilayer graphene \cite{andrei}; and strains between two identical layers in a bilayer \cite{bilayer}. The spatially-dependent van der Waals interaction between the two layers creates a potential energy for electrons in each layer with the periodicity of the moir\'e. Electron hopping between the layers is also modulated with the same spatial periodicity. These two effects are responsible for a number of new electronic phenomena observed in recent years in van der Waals heterostructures, including: the observation of superlattice Dirac points for graphene on BN \cite{hofstader}; collective electronic phases in twisted bilayers and twisted double bilayers \cite{correlations1,correlations2,correlations3,correlations4,correlations5} , and trapping of excitons in the moir\'e potential \cite{moire1,moire2,moire3,moire4}. In these applications of moir\'e engineered bandstructures, the moir\'e potential is of the order of tens of millielectronvolts for typical materials, determined by the typical spacing between the two sheets which is a few angstroms. 

As we explore moir\'e engineering, an open question is whether we can use moir\'e potentials to achieve strong trapping potentials for electrons. In general, large changes in electronic structure are caused when there are significant rearrangements to atomic positions within each layer. The high rigidity of in-plane bonds in van der Waals materials suppresses large atomic rearrangements, especially in the limit where the lattice mismatch is large and the moir\'e wavelength is small. Large effects on in-plane atomic structure are instead found in the limit where the mismatch between the lattice vectors is extremely small. In this regime, atoms try to adjust to the new landscape, forming commensurate regions at the expense of elastic energy. These commensurate regions are separated by topological solitons \cite{sci_rep}, which, in the case of crystalline membranes, are also corrugated in the vertical direction, akin to edge dislocations in solids (these are also called misfit dislocations). The physics of this problem \cite{Pokrovsky} has strong analogies with vortex pinning in type II superconductors \cite{sc}, the onset of charge-density-wave instabilities in metallic transition-metal dichalcogenides \cite{cdw} and atom adsorbates on the surface of graphite \cite{graphite}. As an example of the van der Waals case, consider bilayer graphene. The low energy configuration for a bilayer is bernal stacking, where a carbon atom in one layer sits in the hollow site above the other layer. When a small lattice mismatch is created between the two layers by rotation or heterostrain, this necessarily implies that some regions of the bilayer will have a stacking configuration where one carbon atom sits on top of another, which is energetically unfavorable. In the limit of long wavelength modulations, the system prefers to form large regions of bernal stacking, separated by sharp domain walls or \textit{strain solitons} where a phase slip exists \cite{bilayer}. A lattice of such solitons has been observed in very small angle twisted bilayer graphene  \cite{kerelsky}. Deterministic control over the stacking angle required to obtain uniform soliton networks (precision of 0.01 degree) however remains an unsolved problem. Random networks of solitons have also been observed previously in exfoliated and grown bilayer graphene, and it has been hypothesized that small strain fields present between the layers is responsible for their presence. In this case, the occurrence of the strain solitons is accidental. Being able to create these soliton networks on demand opens up new possibilities for electron trapping and steering. In particular, tightly trapped electrons in semiconductors form the basis of quantum information applications in the solid state. Achieving such trapping via the use of moir\'e engineering provides an alternative pathway to such applications.

In this work, we use strain engineering to create on-demand soliton networks in transition metal dichalcogenides (TMD). In these materials, tensions and curvature not only induce fictitious magnetic fields \cite{science,strain1}, but also modulate the electronic band gap through the deformation potential \cite{exp,paco}. Strain has also been suggested as a possible mechanism to produce single photon emitters of relevance for quantum information technologies \cite{SPE}. Our strain engineering is conceptually explained in Fig.~1a. Consider the two layers of a bilayer TMD. When the two layers are held at zero strain, the layers will be in a commensurate phase. When large tensile strain is applied to only the bottom layer and the top layer is unstrained, the two lattice constants are different from each other resulting in an incommensurate phase. However, as the coupling between layers is not negligible, in the intermediate case of not too large heterostrain between the layers, the top layer will form regions of commensurate stacking that are separated by strain solitons as illustrated in the figure. In our experiments, we choose scanning tunneling microscopy (STM) as a probe that is capable of both structural and spectroscopic interrogation of the solitons. STM, as a surface sensitive probe, requires clean and vibrationally stable samples for high quality imaging and spectroscopy measurements. The creation of soliton networks on demand requires the ability to tune the strain between two adjacent layers of the TMD controllably while imaging the result of the applied strain over a single region of the sample. In our work we overcome these experimental challenges by developing a new apparatus with which we can apply controllable, large uniaxial strain to quantum materials while simultaneously performing STM measurements.

 In order to achieve high strains (above 1\%) in-situ while performing STM measurements, we have constructed a novel piezoelectric device that is compatible with flag-type STM sample holders (Fig.~1b). This device is based on similar designs used recently for transport measurements under high uniaxial strain \cite{wieteska}. In brief, a sample is bridged between two stacks of piezoelectric actuators that form the two ends of a movable bridge. By the application of a voltage to the piezoelectric device, the two ends of the bridge can be moved further apart or closer together, creating uniaxial strain in the sample. Our experiments were carried out on bulk crystals of 2H-MoSe$_2$ that are glued firmly onto ruby substrates that form the ends of the strain bridge, giving the structural rigidity necessary for STM measurements.  A close-up image of a sample affixed to the strain bridge is shown in Fig.~1c. The resting separation of the ends of the bridge was 100 $\mu$m with a maximum variance of $\pm3$ $\mu$m under the application of a voltage (maximum uniaxial strain of $\pm3$ $\%$). Following the construction process, the sample was mechanically exfoliated in UHV conditions immediately prior to the experiment, to avoid exposure to ambient. The crystal is secured at its edges and at the bottom via the epoxy, and in this situation the entire crystal is strained uniformly by application of a voltage to the piezoelectric. However, it is often found that the exfoliation process results in several micron-sized flakes of monolayers that are not bound to the edges of the sample, as illustrated in Fig.~1d. In this situation, the top layer only feels a van der Waals force to the layer below it, and is not mechanically clamped to the piezoelectric apparatus. Thus, the experiment mimics the cartoons of Fig.~1c, where the bottom layer of a bilayer is clamped and strained while the top layer is free to relax its structure as desired. 

We begin our experiments by optically positioning the scanning tip in the gap between the two piezoelectrics. Our experiments are conducted in a homemade STM at a temperature of 77 K, which is sufficient to avoid carrier freeze-out in the semiconducting crystals. We scan across the surface of the crystal till we find a monolayer on the top surface that is not attached to the lateral ends of the crystal as shown in Fig.~1d (Figure~S1 in the Supplementary Material shows the topographic scan across the edge of one such monolayer). We then scan a large area ($1.5 \times 1.5  \mu m^2$) of this top layer far ($> 1  \mu m$) from any edge. The resultant topographic scan is shown in Fig.~1e, obtained at zero strain. This topograph is flat to within about 1 \AA\, with the residual roughness due to the presence of point defects in the crystal \cite{edelberg}. Upon increasing the strain beyond a critical value (determined to be approximately $1.5 \%$), clear signs of the formation of strain solitons are seen in STM topographs. The density of the solitons could be controlled with applied strain. Two such images beyond the critical strain are shown in Fig.~1f and g, at strains of  2\% and 3\% respectively. All three images 1e-g were obtained on exactly the same region of the sample. The process of soliton formation was reversible - by removal of the strain, the solitons could be made to disappear.  

We next consider the atomic scale structure of a single soliton, as shown in Fig.~2a. Each bright spot in this topograph corresponds to a Se atom on the surface of the top layer. The long axis of the soliton coincides with an armchair direction of the crystal. By taking a topographic line cut along the short zig-zag axis, shown in Fig.~2b, we could extract the height profile for the corrugation. The maximum apparent height was measured to be 3 \AA\, with a lateral width of 3.3 nm. We can use the distance between adjacent peaks and troughs in atomically resolved topographs to calculate the local lattice constant across the soliton, as shown in Figure~2c. Here, the blue curve is the local lattice constant across a soliton, while the red curve is the lattice constant across a region without a soliton, both taken from Figure~2d. In this figure, the x(y) direction is the fast (slow) scan direction. The cuts have been chosen at the same y positions, to minimize any scanner or drift-related errors. In the line cut without the soliton, the average lattice spacing is roughly 3.4 {\AA}, consistent with a MoSe$_2$ layer under tensile strain. The lattice spacing in the soliton region is roughly 3.05 {\AA}, consistent with the fact that the soliton relieves tensile strain. The integrated difference in lattice spacing across the soliton is found to be near one lattice constant, indicating that each soliton corresponds to a 2$\pi$ phase slip. We can further confirm this by counting atomic rows parallel to the soliton in the blue and red regions of Fig.~2d. Shown in Fig.~2e are side-by-side comparisons of the topographic excerpts from the two regions. The region across the soliton contains exactly one extra atomic row, implying that the corrugation is a discommensuration of exactly one lattice spacing along a zig-zag axis.
 
The preferential direction of solitons separating different commensurate areas can be understood from the symmetry of the adhesion potential shown in Fig.~2f (details of the calculation can be found in Supplementary Material section S2). The minimum corresponds to the hexagonal (2H) stacking of the bulk material, while the maximum corresponds to a configuration where Se atoms lie on top of each other (XX); the configuration with transition-metal sublattices sitting on top of one another (MM) is a local minimum, close in energy to a saddle point (the stacking at the center of the soliton). Far away from the soliton boundary, the system is in the 2H stacking configuration (darkest areas in Fig.~2f). Going across a single soliton therefore corresponds to moving from one 2H potential minimum to a nearby one.  If this is achieved by moving along the armchair direction (vertical arrow in Fig.~2f), one necessarily has to transition through the XX stacking configuration (brightest areas in Fig.~2f) which has the maximum energy cost. On the other hand, moving along the zigzag direction (horizontal arrow in Fig.~2f) implies that the XX stacking configuration can be successfully avoided, resulting in a lower energy cost. This is also borne out in our finite-element method simulations for soliton formation shown in Fig.~2g-h.  Tensile heterostrain that is applied along the zig-zag direction results in the formation of solitons that are perpendicular to the applied strain. On the other hand, for tensile heterostrain applied along the armchair direction, the resulting solitons are still formed perpendicular to the zig-zag directions. The total strain is then relieved via its components along the zigzag directions. Details of the simulations are provided in the Supplementary Material.
 
The fact that the distance between the first observed solitons is comparable to the size of the sample suggests that the commensurate-incommensurate transition is continuous \cite{Pokrovsky}. Neglecting thermal fluctuations, the critical strain above which strain solitons first appear can be estimated as (see details of the model in the Supplementary Material section S3) \begin{equation}
\bar{u}_c=\frac{\eta_{\varphi}8}{\pi}\sqrt{\frac{V}{\lambda+2\mu}}\,,
 \end{equation}
where $\lambda,\mu\approx 3$ eV/\AA$^2$, are the Lam\'e coefficients of the monolayer \cite{elastic_modulus}, $\eta_{\varphi}$ is a numerical factor depending on the orientation of the sample ($\eta_{0}=1$ if the tension is along a zig-zag axis, $\eta_{\pi/6}\approx1.22$ in the case of an armchair direction) and $V\approx 43$ meV/nm$^2$ is the scale characterizing the free-energy difference between different stacking configurations. The width of the solitons is independent of the amount of strain and goes like \begin{equation}
\ell=\frac{\tilde{a}}{4\pi}\sqrt{\frac{\lambda+2\mu}{V}}\approx 4\, \textrm{nm},
\end{equation}
with $\tilde{a}\approx3.4$ \AA\, being the lattice constant of strained MoSe$_2$. The model tends to overestimate this length scale because it does not take into account the corrugation. Note also that at finite temperature, the critical strain for the appearance of solitons should be a bit smaller than the estimate of Eq.~(1) (which gives $\bar{u}_c\sim$ 2\% for a tension along the armchair direction) \cite{Pokrovsky}. Moreover, both thermal fluctuations and pining forces created by disorder in the sample break the positional long-range order of the solitons in the incommensurate state. Additionally, the three-fold symmetry of the adhesion potential (only weakly broken by the uniaxial strain) implies the existence of more than one easy direction for the solitons. Despite the free-energy cost of soliton crossings (see below), these two-dimensional structures carry a lot of entropy, so even at low temperatures fluctuations can contribute to stabilize them.

We next consider the formation and properties of soliton crossings. Figure~3a-c shows a sequence of STM images over the same region of the sample at strains of 1.8, 2.0 and 2.2 \% respectively. Fig.~3a shows how a new soliton nucleates at the core of a dislocation (the endpoint of a soliton, see also Fig.~S2 in the Supplementary Material). Upon increasing the strain slightly (Fig.~3b), the soliton propagates through the area, and we see the formation of soliton crossing, where discommensurations along two zig-zag directions coincide. Upon further increasing the strain (Fig.~3c), the soliton crossing is seen to split into two separate vertices where three solitons meet. Far away from the soliton, the stacking order for the top bilayer is 2H. On the solitons themselves, the stacking order evolves continuously. A consideration of how the stacking order evolves near a soliton crossing helps us understand the splitting of a single soliton crossing into two three-fold vertices. Shown in Fig.~3d is a schematic of the soliton crossing of Fig.~3b. Each of the solitons corresponds to a phase slip of one lattice constant along the green and blue arrow directions respectively. The path of the system in stacking order configuration space is visualized by arrows of the same colors in Fig.~3e. The soliton crossing itself corresponds to the shaded region in Fig.~3d. We can clearly see from this figure that the soliton crossing corresponds to a region where all stackings (shaded area in Fig.~3e) are sampled, resulting in a high energy cost. This soliton crossing is therefore unstable, and relaxes quickly as shown in Fig.~3c into two separate vertices connected by a third soliton. The schematic for this image and the corresponding paths in configuration space are shown in Fig.~3f-g. The formation of the honeycomb-like soliton network or a multi-domain stripe phase as seen in Fig.~1 is a consequence of these relaxation processes.

The two separate vertices have rather different properties. In order to understand the difference between the two, consider the configuration of the system as one traverses along the soliton across each of the two vertices (blue and orange dashed lines in Fig.~3f). The corresponding path in configuration space is shown by the dashed lines in Fig.~3g. We can clearly see that for one of the intersections (blue dashed line), going across the soliton vertex results in the system going through the XX stacking configuration, while for the other soliton vertex the system goes through the MM stacking configuration. Henceforth, we term these the XX and MM vertices, respectively. In order to further confirm this scenario, we imaged several of these shorter solitons with STM with high resolution; one representative example is shown in Fig.~3h. The two vertices showed a distinct topographic contrast (one brighter or higher than the other), which was systematically reproduced over the sample. It is also seen that the brighter vertices are a sharper topographical feature than the darker one, which occupies a larger spatial size. Figures~3i-j show atomic resolution taken at each vertex. Figure~3i reveals a triangular pattern at the bright vertex, whose apparent height decays fast on the atomic scale. This suggests that Se atoms are lifted by the local XX stacking in this region, and therefore are lifted with respect to the surrounding commensurate regions. In contrast, at the darker vertex, panel~j, the topographic contrast resembles two interpenetrating triangular lattices, resulting in a honeycomb geometry. A plausible reason for this is that the Mo atoms in this region are pushed upward by the local MM stacking, allowing them to be visible in the topograph. This identification of the two vertices based on local topography is also compatible with our previous observation that the darker vertices are more extended in space: the adhesion potential is flatter around the MM stacking, so the system explores other configurations close in energy.

Having identified the two distinct vertices in the honeycomb soliton network, we now consider the consequences of these soliton vertices for the electronic structure of the material, using scanning tunneling spectroscopy (STS) measurements of the local density of states. Figure~4a shows STS spectra averaged on the commensurate 2H regions of the material (far from the solitons) as a function of strain. The gap decreases with increasing strain (see values in Tab.~S1 of Supplementary Material); to the best of our knowledge, this is the first direct measurement of the single-particle band gap reduction by tensile strain in this class of materials. We now consider the spectrum in the soliton vertex. In Fig.~4b and its inset, we compare the spectra recorded on a commensurate region (2H), a dark soliton vertex (MM), and a bright vertex (XX) at a strain of 2.5\%. The most prominent feature is the presence of electronic states deep inside the gap at 75 and 175 meV in the XX vertex (main panel of Fig.~4b), which are absent for both the commensurate 2H regions as well as the MM soliton vertex (the inset shows zoomed out versions of the spectra). States at similar energies are observed in all XX vertices. We note that the width of the in-gap resonances in our spectra is large due to the relatively high temperature of our measurement. To confirm the spatially bound nature of the in-gap states, we perform spectroscopic imaging at the energy of the stronger resonance, the results of which are shown in Fig.~4c. In this image, the color scale depicts the magnitude of the local density of states, while the simultaneously acquired topography is rendered in the three dimensional height map. It is clearly seen that the in-gap resonance is confined solely to the XX soliton vertex that has the higher topographic height, while it is absent on the MM vertex. The spatial width of the resonance is about 2 nm. Besides the in-gap states, weaker resonances are present in both soliton vertices in the semiconducting bands, starting at the band edges (better resolved in the conduction band than in the valence band). To show the nature of these band resonances, we perform an angular average of spectra around the XX site, and display the angular-averaged spectra as a function of distance from the XX site in Fig.~4d for a range of energies.  The conduction band edge is indicated by the arrow on this image. The resonances within the conduction band are equally spaced by about 75 meV, and their spatial structure resembles the wave functions of a two-dimensional harmonic oscillator, with alternating resonances displaying opposite parities. This sequence of energy levels with alternating parity is compatible with strain-induced bound states recently discussed in the context of TMD bubbles \cite{paco}. These features are also present in STS maps around MM crossings but much more attenuated in that case. The reason is that these crossings are wider, their corrugation is therefore smaller, resulting in a weaker confining potential and diminished resonances with respect to the XX crossings. 

Among the possible mechanisms for electronic confinement of the states in the semiconductor band, two are the most relevant in this case: non-uniform strains and spatially-modulated interlayer couplings due to changes in the atomic registry. Apart from strain-induced pseudo-magnetic fields, local changes in the deformation potential (the latter responsible for the modulation of the band gap) can also give rise to confinement in this case \cite{paco}. The coupling between layers affects both the interlayer tunneling, which determines the layer splitting of the bands, and the crystal fields and intralayer hopping, which control the position of the band edges. This latter effect, along with the deformation potential, contributes to create a smooth (on the scale of the microscopic lattice) confining potential within the solitons. Its microscopic origin is thus the same as the moir\'e superlattice potentially created by lattice mismatch in TMD heterobilayers, where the same kind of features in the local density of states \cite{heterobilayer1,heterobilayer2} are observed.

The two in-gap states, however, do not belong to this sequence of odd-even parity associated with the smooth confining potential. Instead, they are electronic states that are deeply confined within the bandgap as shown above. The fact that these resonances appear deep inside the gap points to a sharper perturbation on the atomic scale, similar to the presence of bound states around chemical dopants in semiconductors. Figure~4e illustrates a simple model for this scenario, in which these states originate from multiple scattering off a sharp \textit{stacking defect} at XX soliton crossings. In the TMD semiconductors, conduction and valence bands are dominated by orbitals localized on the metal sites (in red in the figure). The effective hopping between these orbitals (represented by dashed green arrows) are assisted by Se orbitals (localized on blue sites in the figure). At the XX soliton vertex, the strong coupling between Se atoms sitting on top of each other can lead to the passivation of their orbitals, interrupting these virtual hopping processes and mimicking the effect of a Mo vacancy. Scattering off such a defect can transfer spectral weight from the electron-hole continuum to build resonances within the band gap. To model this quantitatively, we use a simple tight-binding model with hopping parameters for monolayer MoSe$_2$ \cite{TB_model} (details in Supplementary Material section S6). We then passivate bonds around a single Mo atom by assuming an infinite energy on the central Mo site (\textit{strong scatterer} limit), and calculate the local density of states. This model shows generically the presence of two deeply trapped states.  Figure~4~d shows the local density of states calculated at different distances from the stacking defect. The perturbations to the local density of states disappear within about two lattice constants from the central Mo site in this simple model. While the true experimental situation is undoubtedly more complex, the basic features that there are two in-gap states that are localized well within the soliton vertex (full width at half maximum of 1.5 nm) are captured well by the simple model.

In conclusion, we have demonstrated a commensurate-incommensurate transition driven by uniaxial strain in a Van der Waals material. In the incommensurate state, tensile strain is relieved through the formation of a honeycomb-like soliton network. Local stacking at the soliton crossings correspond to extreme configurations of the adhesion potential, in which either chalcogen or transition-metal atoms sit on top of one another. The morphology of the soliton network is ultimately linked to the symmetry of the interlayer potential; honeycomb-like soliton networks are also observed in moir\'e systems like graphene on hexagonal boron nitride \cite{gr/hBN} or twisted bilayers of transition-metal dichalcogenides \cite{twisted,prl}, for example. Solitons confine electrons; this effect is stronger at XX stacked crossings, where the enhanced coupling between chalcogen atoms gives rise to midgap states. This system is a new possible avenue for many-body effects provided the strong electron correlation intrinsic to transition-metal compounds, combined with spatial confinement and tunability by strain engineering.

\section*{Acknowledgments}
This work is supported by the National Science Foundation via grant DMR-1610110 (DE) and by the NSF MRSEC program through Columbia in the Center for Precision Assembly of Superstratic and Superatomic Solids (DMR-1420634, HO and ANP). Support for STM measurements is provided by the Air Force Office of Scientific Research via grant FA9550-16-1-0601. The computational work is supported primarily by contract W911NF-16-1-0447 from the Army Research Office (V.B.S.) and by grant CMMI-1727717 (H.K.) from the U.S. National Science Foundation. 

\section*{Supplementary materials}
Supplementary Text\\
Figs. S1 to S3\\
Table S1\\

\clearpage

\section*{Figure Captions}

\noindent {\bf Fig.~1.} Strain soliton networks in MoSe$_2$. a) Schematic of the commensurate, partially incommensurate (strain solitons), and totally incommensurate bilayer structures. b) Piezoelectric based device for performing STM measurements under large uniaxial strain. c) Optical image of a MoSe$_2$ crystal affixed across the two independent piezoelectric stacks. The sample is bounded to each stack by epoxy. d) Schematic of a free monolayer on the top surface of the crystal after cleaving in vacuum. Strain was transmitted only through the bulk material underneath. e)-g) STM topographic images of a 1.5 $\times$1.5 $\mu$m$^2$ region of the top monolayer at strains of 0\%, 2\% and 3\% respectively (V=0.8 V, I=100 pA). The sample was first flat (residual height variations due to crystal defects). Above a critical strain of 1.5 \% solitons begun to form, increasing in density as strain grew.
\\
\\
\noindent {\bf Fig.~2.} Atomic structure of a single soliton. a) Atomically resolved STM topographic image (V=1 V, I=100 pA) of one of the solitons where the top lattice of Se atoms is well resolved. b) Height profile across the soliton, extracted from the previous image. c) Local lattice spacing (projected on the plane) across the red and blue regions highlighted in panel d). In the soliton region, the local lattice spacing is reduced, corresponding to a release of the tensile strain in the top layer. e) Excerpts from panel d) showing atomic resolution on (blue) and off (red) a soliton. A count of atoms across the soliton contains one extra atomic row, indicating a discommensuration of one lattice spacing across the soliton. f) Free-energy landscape of stacking configurations (details of the calculation in the Supplementary Material section S2). g)-h) Soliton solutions deduced from our numerical simulations when strain is applied along the zig-zag (g) and armchair (h) directions, respectively.
\\
\\
\noindent {\bf Fig.~3.} Soliton crossings and vertices. a)-c) STM topographic images (V=1V, I=50 pA) at strains of 1.8\%, 2.0\% and 2.2\% respectively, showing the evolution from a) a pinned soliton endpoint to b) formation of a second soliton with a soliton crossing to c) splitting of the soliton crossing into two vertices. d) Schematic of the soliton crossing in b). e) Corresponding path of the system in configuration space in the various regions. Across each of the solitons, the system transits from one 2H minimum to a neighboring one as indicated by the green and blue colored arrows. In the region of the crossing, the system explores all possible stacking configurations, including XX and MM extremes. f) schematic of the two vertices in c). g) Corresponding path of the system in configuration space.  Paths (blue and orange dashed lines) connecting commensurate regions with the third soliton (red arrow) indicate how the two vertices correspond to XX and MM stackings. h) STM topographic image (V=1V, I=50 pA) of a short soliton connecting to two vertices with XX and MM stacking configurations respectively. i)-j) STM topographic images (V=1V, I=50 pA) of the XX and MM soliton vertices respectively. For the XX soliton in i), the topographic height decays quickly on the atomic scale. For the MM soliton vertex in j), the triangular lattice is blurred on a longer length scale, replaced by a honeycomb pattern at the center of the crossing.
\\
\\
\noindent {\bf Fig.~4.} Spectroscopic properties of soliton vertices a) STS spectra recorded on bulk MoSe$_2$. The indirect gap closes as the strain grows. b) Local density of states within the semiconducting gap recorded on a XX soliton crossing (solid line) and commensurate 2H region (dashed line) at a strain of 2.5\%. The XX vertices display clear midgap states. The inset shows the LDOS on larger energy scales, also showing that the MM soliton crossing does not display midgap states. Additionally, the spectra at both vertices display smaller resonances in the semiconductor bands, more clearly resolved in the conduction band. c) Angular-averaged STS maps in the conduction band around a XX crossing. The conduction band edge is denoted with the arrow on the y-axis. The resonances in the conduction band have alternating odd-even parity as expected for the harmonic oscillator. The two midgap states do not follow this sequence d) Model for a \textit{stacking defect} on a XX soliton vertex: the passivation of the Se (in blue) orbitals due to a strong inter-layer coupling interrupts the effective hopping between Mo (in red) orbitals, which dominate the low energy bands, leading to an effective vacancy. e) Calculated density of states at different distances from the stacking defect for a monolayer (details of the model and the calculation in the Supplementary Material section S6).

\clearpage

\begin{figure}[h!]

\includegraphics[width=1.\textwidth]{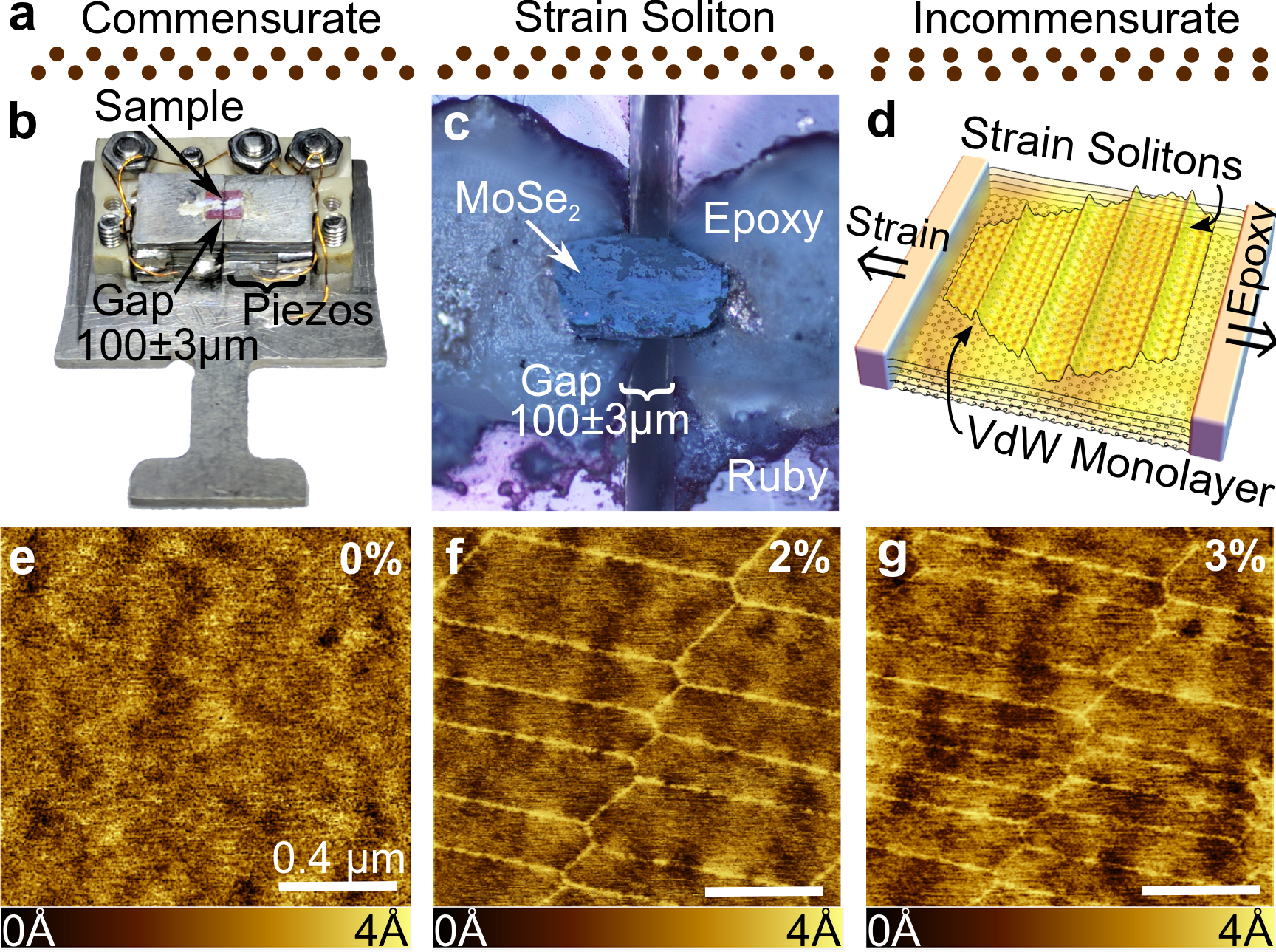}
 
\end{figure}
\noindent {\bf Figure 1} 

\clearpage

\begin{figure}[h!]

\includegraphics[width=1.\columnwidth]{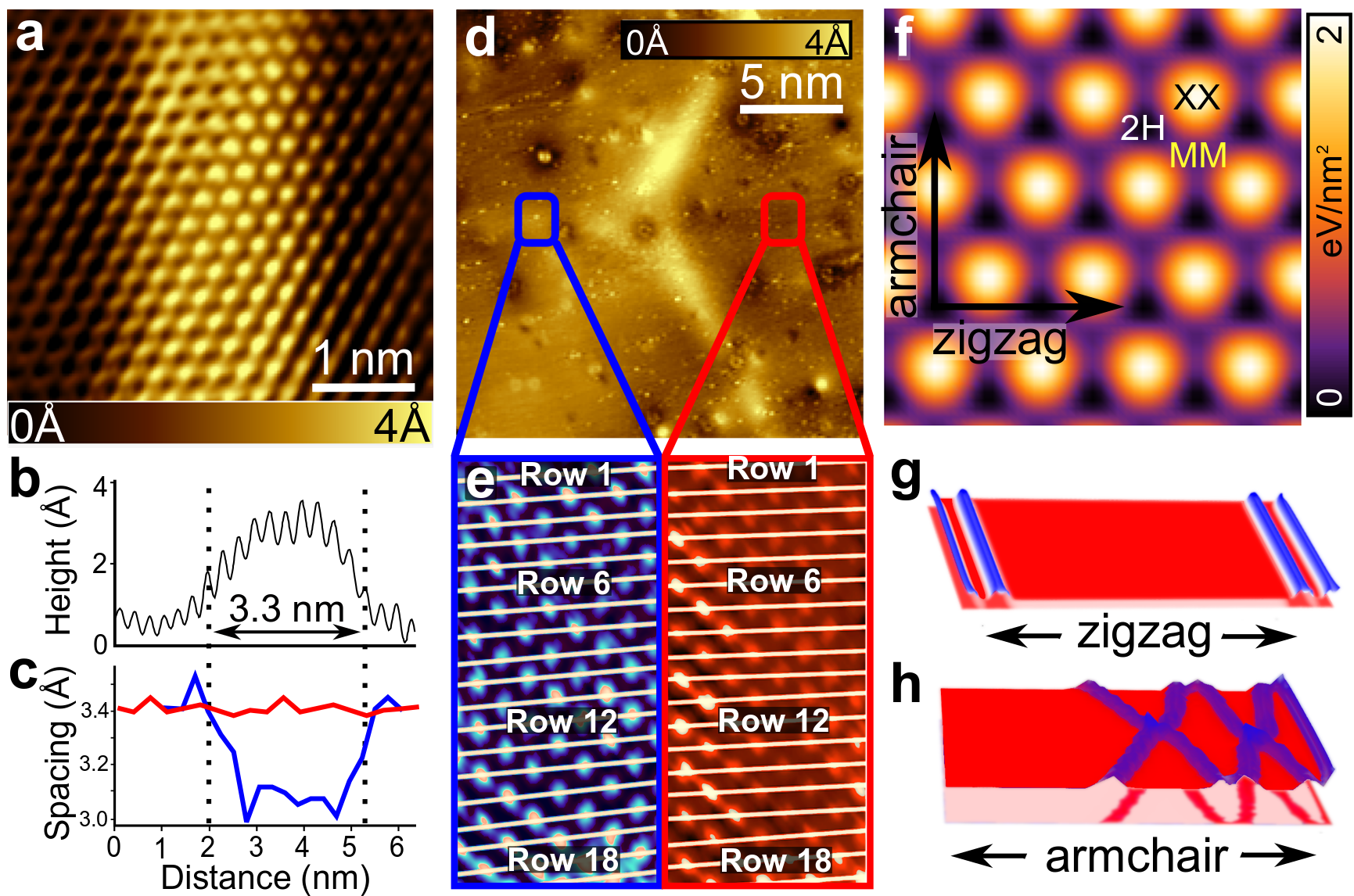}

\end{figure}
\noindent {\bf Figure 2} 

\clearpage

\begin{figure}[h!]

\includegraphics[width=1\textwidth]{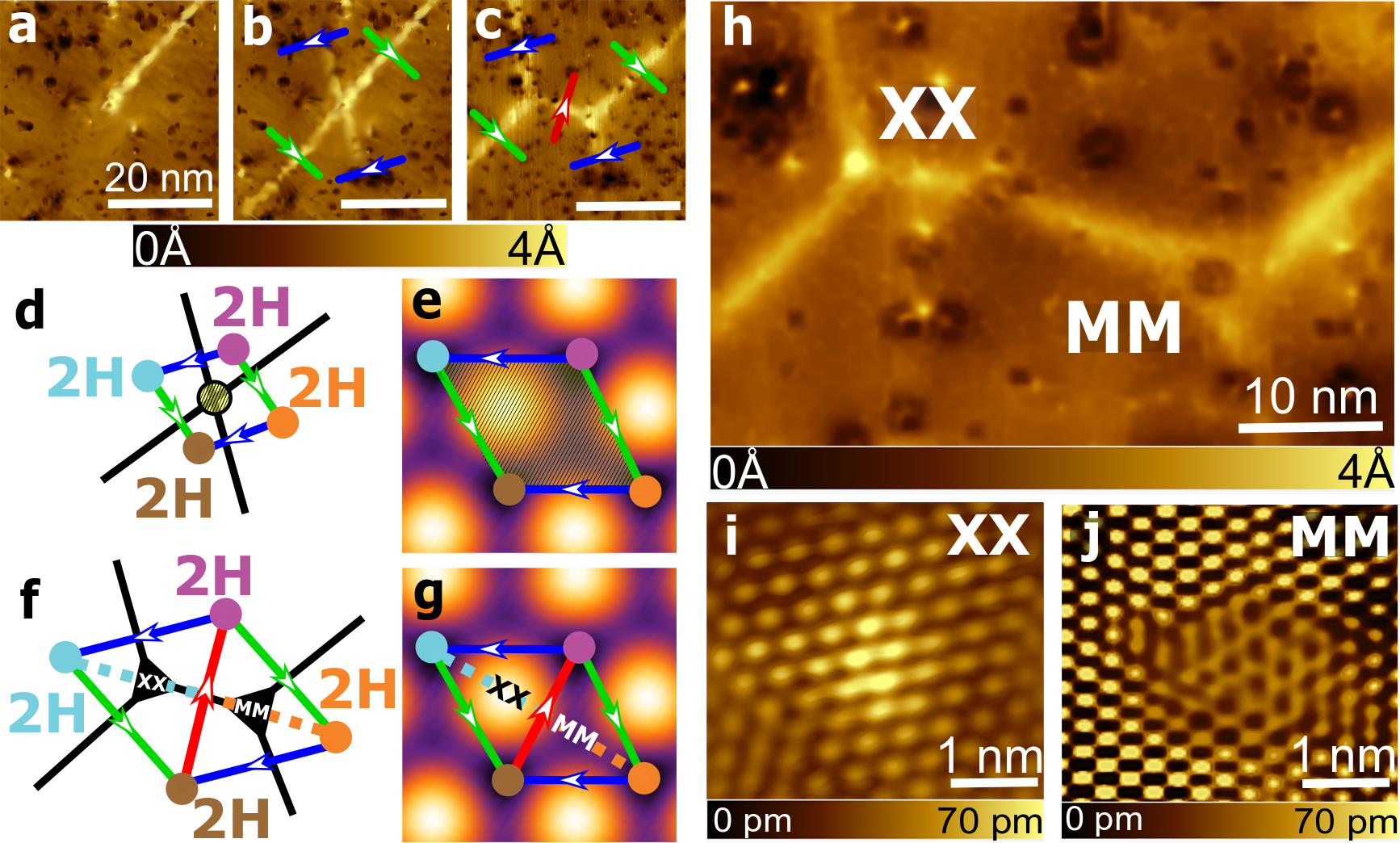}

\end{figure}
\noindent {\bf Figure 3} 

\clearpage

\begin{figure}[h!]

\includegraphics[width=1\textwidth]{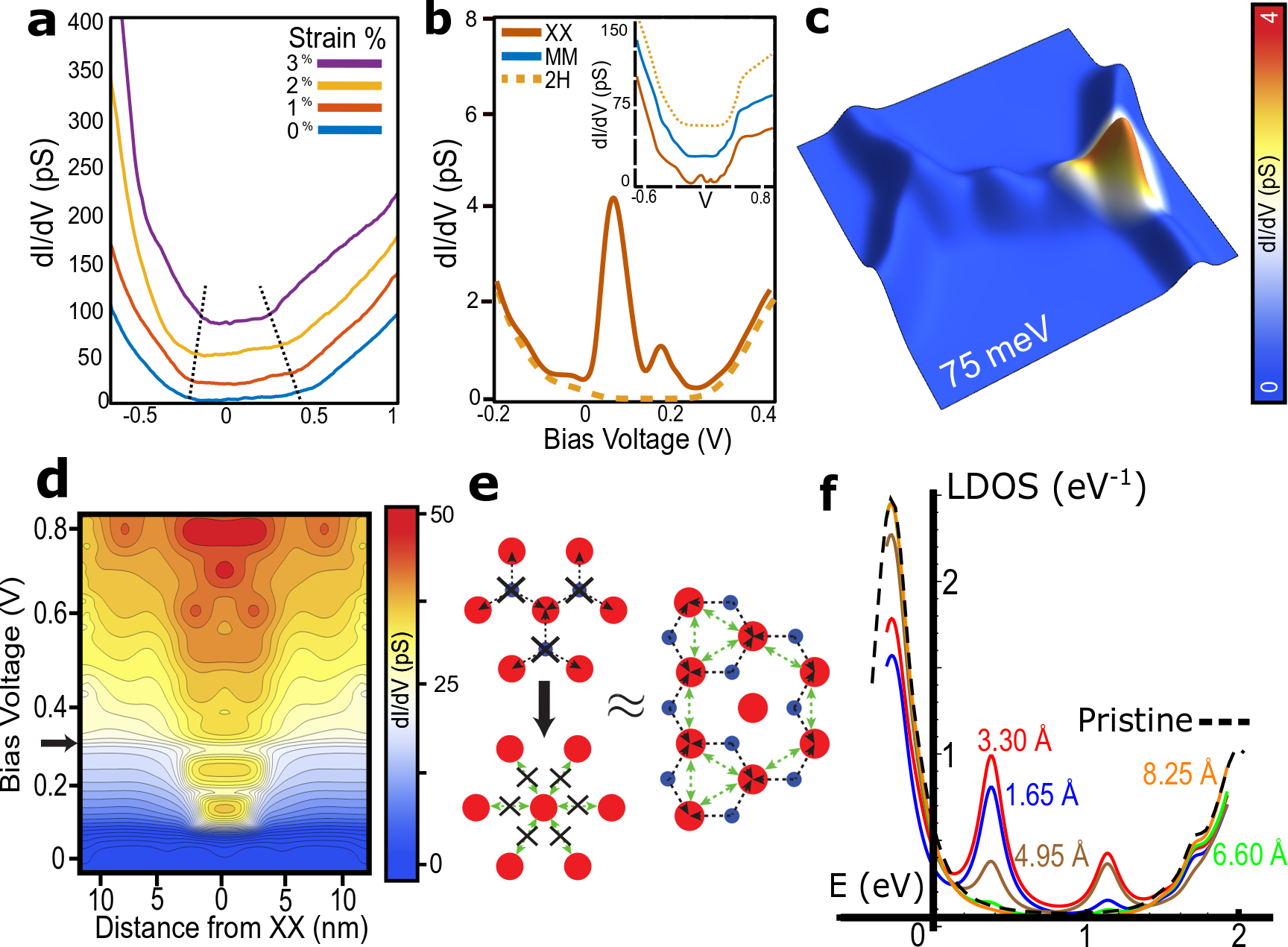}

\end{figure}
\noindent {\bf Figure 4}

\newpage

\section*{Supplementary Information} 

\subsection*{S1: Soliton termination on monolayer edge}

To prove that solitons were occurring in a single layer, we imaged with STM the monolayer edge. Figure~S1 shows how solitons terminated at the boundary of the free monolayer and did not propagate into the bulk. We extracted the height profile across the edge, consistent with a single-layer step in MoSe$_2$.

\subsection*{S2: Adhesion potential}

To compute the interlayer interaction energy, we have used first principles calculations based on density functional theory (DFT) as implemented in the Vienna ab-initio simulation (VASP) code \cite{DFT1}. The exchange-correlation functional was included within the Perdew-Burke-Ernzerhof (PBE) generalized gradient approximations (GGA), and a plane-wave expansion of the wave function was performed with an energy cutoff of $400$ eV. To sample the first Brillouin zone a $\Gamma$-centered k-point mesh of $24\times24\times1$ was adopted. The long-range interlayer van der Waals interactions were included through the DFT-TS scheme \cite{DFT2}. In order to estimate the interlayer interaction energy as a function of atomic registry, single point energies were recorded for different stacking configurations, and fitted to a first-star expansion of the form\begin{equation}
V_{ad}\left(\mathbf{d}\right)=6V-2V\sum_{i=1}^3\cos\left(\mathbf{G}_i\cdot\mathbf{d}\right)-2V'\sum_{i=1}^3\sin\left(\mathbf{G}_i\cdot\mathbf{d}\right),
\end{equation}
where $\mathbf{G}_{1,2}$ are primitive vectors of the MoSe$_2$ reciprocal lattice, $\mathbf{G}_3=-\mathbf{G}_1-\mathbf{G}_2$, and $\mathbf{d}$ represents a relative displacement of one layer with respect to the other. The values of the coefficients are $V=0.136$ eV/nm$^2$, $V'=0.180$ eV/nm$^2$, slightly larger than in \cite{32}.

\subsection*{S3: Model for soliton formation}

In order to estimate the critical strain at which solitons first appear, we consider the following simplified model for the mechanical free energy of the top MoSe$_2$ monolayer:
\begin{equation}
\mathcal{F}\left[\mathbf{u}\left(\mathbf{r}\right)\right]=\mathcal{F}_{el}\left[\mathbf{u}\left(\mathbf{r}\right)\right]+\mathcal{V}_{ad}\left[\mathbf{u}\left(\mathbf{r}\right)\right].
\end{equation}
The first term accounts for the elastic energy due to displacements $\mathbf{u}(\mathbf{r})$ with respect to crystallographic positions, and can be expressed in terms of the components of the strain tensor $u_{ij}=(\partial_i u_j+\partial_j u_i)/2$ as\begin{equation}
\mathcal{F}_{el}\left[\mathbf{u}\left(\mathbf{r}\right),h\left(\mathbf{r}\right)\right]=\frac{1}{2}\int d^2\mathbf{r}\left[\lambda \left(u_{ii}\right)^2+2\mu\, u_{ij}u_{ij}\right],
\end{equation}
where repeated indices are summed up and $\lambda$, $\mu$ are the Lam\'e coefficients of a MoSe$_2$ monolayer. In our estimates, we take $\mu,\lambda \approx3$ eV/\AA$^2$. Note that we are neglecting here the contribution from out-of-plane deformations provided that their energy cost is negligible for length scales longer than $\sqrt{\kappa/(\lambda+2\mu)}\sim a$, where $\kappa\approx 1$ eV is the typical value of the bending rigidity and $a=3.3$ \AA\, is the lattice constant of MoSe$_2$. The second term in Eq.~(1) accounts for the adhesion energy computed in the previous section.

We consider the situation in which tensile uniaxial strain $\bar{u}>0$ is applied to the the multilayer along the direction $\mathbf{\hat{t}}$. The displacement field in the free layer can be conveniently parametrized as $\mathbf{u}\left(\mathbf{r}\right)=\bar{u}\left(\mathbf{\hat{t}}\cdot\mathbf{r}\right)\,\mathbf{\hat{t}}+\mathbf{d}\left(\mathbf{r}\right)$. The advantage of this parametrization is that the strain $\bar{u}$ enters now as a boundary term, affecting the energetics but not the equilibrium equations for the relative displacement or \textit{discommensuration} field $\mathbf{d}(\mathbf{r})$,\begin{equation}
\left(\lambda+\mu\right)\nabla\left(\nabla\cdot\mathbf{d}\right)+\mu\nabla^2\mathbf{d}=\frac{\partial V_{ad}}{\partial\mathbf{d}},
\end{equation}
where $V_{ad}$ represents the adhesion-energy density in Eq.~(1). We solved these equations numerically in a rectangular domain with strain-free boundary conditions \cite{20}. In the experiments (and the analytical solution below) the repulsive interaction between solitons creates a uniform distribution of them across the sample. Finite-size effects in our simulation attenuates these forces, resulting in the accumulation of solitons near the edges shown in panels g and h of Fig.~2.

In order to estimate the critical strain, we look for analytical solutions of a simplified one-dimensional problem in which the possible configurations are restricted to $\mathbf{d}(\mathbf{r})=\phi(\xi)\mathbf{\hat{u}}_{zz}$; here $\mathbf{\hat{u}}_{zz}$ is a unit vector along a zig-zag direction, for which the discommensuration is most likely to occur; in this approximation, the adhesion potential simplifies to\begin{equation}
V_{ad}\big(\phi\big)=8V\sin^2\left(\frac{\phi}{\tilde{a}}\right).
\end{equation} The coordinate $\xi$ is defined, in principle, along an arbitrary axis $\mathbf{\hat{u}}_{\xi}$. A simple inspection of Eq.~(4) reveals that both tensile ($\mathbf{\hat{u}}_{\xi}\parallel \mathbf{\hat{u}}_{zz}$) and shear ($\mathbf{\hat{u}}_{\xi}\perp\mathbf{\hat{u}}_{zz}$) one-dimensional solutions exist. Nevertheless, shear solitons are always more energetic and the transition to an incommensurate state is through the formation of tensile solitons, which agrees with the experimental observations. 

Thus, we focus on one-dimensional tensile soliton solutions of the simplified problem by taking $\mathbf{\hat{u}}_{\xi}= \mathbf{\hat{u}}_{zz}$. Solutions with $\phi'\neq0$ can be expressed implicitly in terms of an incomplete elliptical integral of the first kind as\begin{equation}
\frac{\xi-\xi_0}{\ell}=\pm\frac{1}{\sqrt{C}}\,F\left(\frac{\pi\phi\left(\xi\right)}{\tilde{a}}|-\frac{1}{C}\right),
\end{equation}
where $\ell$ is the soliton width given in Eq.~(2) of the main text. The two possible branches correspond to positive/negative discommensurations, but for $\bar{u}>0$ only the latter is sensible. Inverting this last equation (note that the right-hand side is a monotonic function of $\phi$) gives an array of domain walls of width $\ell$ and period \begin{equation}
\mathcal{L}=\frac{\ell}{\sqrt{C}}\, F\left(\pi|-\frac{1}{C}\right),
\end{equation}
parametrized by integration constants $\xi_0$ and $C$. The inverse of $\mathcal{L}$ represents the density of solitons, which can be taken as the order parameter of the commensurate-incommensurate transition. The energy of solutions with $\mathcal{L}^{-1}\neq0$ is independent of the soliton center $\xi_0$ (where $\phi=0$), which represents the Goldstone mode associated with the broken translational symmetry in the incommensurate state (phason). The constant of integration $C$ (and therefore the density of solitons via Eq.~7) can be determined by minimization of the free-energy cost of the soliton solution, which relates $C$ with the amount of strain via\begin{equation}
\bar{u}=\frac{\lambda+2\mu}{\lambda+2\,\mu\,\cos^2\varphi}\times\frac{1}{\pi\ell}\int_0^{\tilde{a}} d\tilde{\phi}\,\sqrt{\frac{V_{ad}\big(\tilde{\phi}\big)}{8V}+C}\,.
\end{equation}
For a given value of the strain, only soliton solutions with $C$ satisfying this relation correspond to a free-energy minimum. In this last expression, $\varphi$ is the angle between the applied tension $(\mathbf{\hat{t}})$ and the director vector of the soliton (along a given zig-zag direction).

Since $\bar{u}$ grows monotonically with $C$, the critical strain above which minimum solutions with $\mathcal{L}^{-1}\neq 0$ first exist is given by the condition $C=0$, defining the critical strain in Eq.~(1) of the main text. Above $\bar{u}_c$, in fact, the free energy of the soliton solution is lower than the energy of the commensurate state, therefore, $\bar{u}_c$ sets the critical strain above which the system undergoes a continuous phase transition to a striped incommensurate state. Close to the transition (from above), our solution resembles a train of sine-Gordon domain walls of width $\ell$, whose separation $\mathcal{L}$ scales with the inverse of the logarithm of $\bar{u}-\bar{u}_c$; specifically, $\mathcal{L}=\ell\ln[4\bar{u}_c/(\bar{u}-\bar{u}_c)]$. Note, however, that thermal fluctuations destroy long-range order in the positions of the solitons, reducing the critical value of the strain given in Eq.~(1) of the main text, which has to be taken as an upper bound. Entropic terms associated with steric repulsion between solitons and not included in the present model alter the scaling relation between the soliton density and the amount of strain; in particular, $\mathcal{L}\sim(\bar{u}-\bar{u}_c)^{-1/2}$ \cite{18}.
 
\subsection*{S4: Soliton dislocations}

Disorder over the sample creates pinning forces on the solitons, breaking the positional order of the incommensurate state. Disorder can also induce an inhomogeneous distribution of strain that favors the formation of dislocations in the stripe ordering. Figure~S3 shows one of these dislocations occurring in a sample strained along a zig-zag direction. Solitons around the dislocation core (the endpoint of the central soliton) adjust to keep their separation as uniform as possible and minimize the free-energy cost of these defects.

\subsection*{S5: Band gap as a function of strain}

Table~S1 shows the lattice constant and electronic band gap recorded on commensurate (2H) areas for different values of the applied strain.

\begin{table}
\centering
\begin{tabular}{|c||c||c|}
\hline
Strain (\%) & Lattice constant (\AA) & Band gap (eV) \\
\hline
\hline
0 & 3.30 & 0.86\\
\hline
1 & 3.33 & 0.73\\
\hline
2 & 3.36 & 0.6\\
\hline
3 & 3.40 & 0.47\\
\hline
\end{tabular}\\
$ $\\
\noindent {\bf Tab. S1}: Band gap of MoSe$_2$ for different values of strain. 
\end{table}

\subsection*{S6: Model for midgap states}

The model consists of two terms, $\hat{H}=\hat{H}_{it}+\hat{V}$, the first one describing itinerant electrons in the first layer, and $\hat{V}$ being the perturbation introduced by the XX stacking defect on a bright soliton crossing. For the former, we adapt the 3-orbital tight-binding description of the bands in monolayer MoSe$_2$ \cite{31}, which includes all possible symmetry-allowed hoppings between $d_{x^2-y^2}$, $d_{xy}$, and $d_{z^2}$ orbitals of Mo atoms, represented in red in panel d of Fig.~4 in the main text. Note that these processes are mediated by direct hoppings to the surrounding Se atoms, represented in blue in the same figure. In this simplified framework, the density of electronic states localized on the first layer is just $\rho\left(E,\mathbf{r}\right) =-$Im Tr $\left\langle \mathbf{r}\right|\hat{G}^R\left(E\right)\left|\mathbf{r}\right\rangle/\pi$, where the retarded Green operator is defined as $\hat{G}^R(E)\equiv[E+i\delta-\hat{H}]^{-1}$ and the trace is in orbital space. Summing up multiple scattering events with the stacking defect, the Green operator can be written as\begin{equation}
\hat{G}^R\left(E,\mathbf{r},\mathbf{r}\right)=\hat{G}_0^R\left(E,\mathbf{r},\mathbf{r}\right)+\sum_{\mathbf{r}_1,\mathbf{r}_2}\hat{G}_0^R\left(E,\mathbf{r},\mathbf{r}_1\right)\cdot\hat{T}\left(E,\mathbf{r}_1,\mathbf{r}_2\right)\cdot\hat{G}_0^R\left(E,\mathbf{r}_2,\mathbf{r}\right),
\end{equation} 
where $\hat{G}_0^R(E)\equiv[E+i\delta-\hat{H}_{it}]^{-1}$ is known and the $T$-matrix is the solution to the equation\begin{equation}
\hat{T}\left(E,\mathbf{r}_1,\mathbf{r}_2\right)=\hat{V}\left(\mathbf{r}_1,\mathbf{r}_2\right)+\int d\mathbf{r}_3\int d\mathbf{r}_4\,\hat{V}\left(\mathbf{r}_1,\mathbf{r}_3\right)\cdot\hat{G}_0^R\left(E,\mathbf{r}_3,\mathbf{r}_4\right)\cdot\hat{T}\left(E,\mathbf{r}_4,\mathbf{r}_2\right),
\end{equation}
where $\hat{V}(\mathbf{r}_1,\mathbf{r}_2)=\langle\mathbf{r}_1|\hat{V}|\mathbf{r}_2\rangle$.

Next, we assume that a strong hybridization between the orbitals at Se atoms of different layers interrupt the virtual hopping between Mo nearest neighbors, effectively creating a \textit{vacancy} in the Mo lattice on a site labelled by $0$. The latter can be simulated by passivating the orbitals with large on-site energies (greek indices label orbital states),\begin{equation}
\hat{V}=\sum_{\mu,\nu}\left[\hat{\mathcal{V}}\right]_{\mu,\nu}\left|0,\mu\right\rangle\left\langle0,\nu\right|,
\end{equation}
thus, $\hat{V}(\mathbf{r}_1,\mathbf{r}_2)=\hat{\mathcal{V}}\,\delta_{\mathbf{r}_1,0}\,\delta_{\mathbf{r}_2,0}$. The solution of Eq.~(9) of the form $\hat{T}\left(E,\mathbf{r}_1,\mathbf{r}_2\right)=\hat{T}\left(E\right)\,\delta_{\mathbf{r}_1,0}\,\delta_{\mathbf{r}_2,0}$, with \begin{equation}
\hat{T}\left(E\right)=\left[1-\hat{\mathcal{V}}\cdot\hat{G}_0^R\left(E,0,0\right)\right]^{-1}\cdot\hat{\mathcal{V}}.
\end{equation}
If the energy of the defect is much larger than the bandwidth, we can take $\hat{T}\left(E\right)\rightarrow -\left[\hat{G}_0^R\left(E,0,0\right)\right]^{-1}$. In this limit, the resonances are completely determined by the \textit{pristine} density of states deduced from $\hat{H}_{it}$. The real (in blue) and imaginary (in red) parts of the (trace of the) $T$-matrix are represented in Fig.~S4, presenting two poles (a doublet and a singlet, no spin-orbit coupling is included) at energies within the band gap. These poles give rise to midgap resonances that decay very quickly on the atomic scale, as shown in the computed local density of states in Fig. 4 of the main text.

\clearpage

\begin{figure}[h!]

\includegraphics[width=0.4\textwidth]{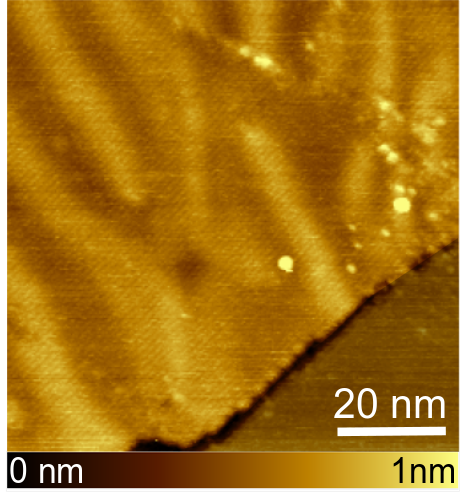}
\includegraphics[width=0.55\textwidth]{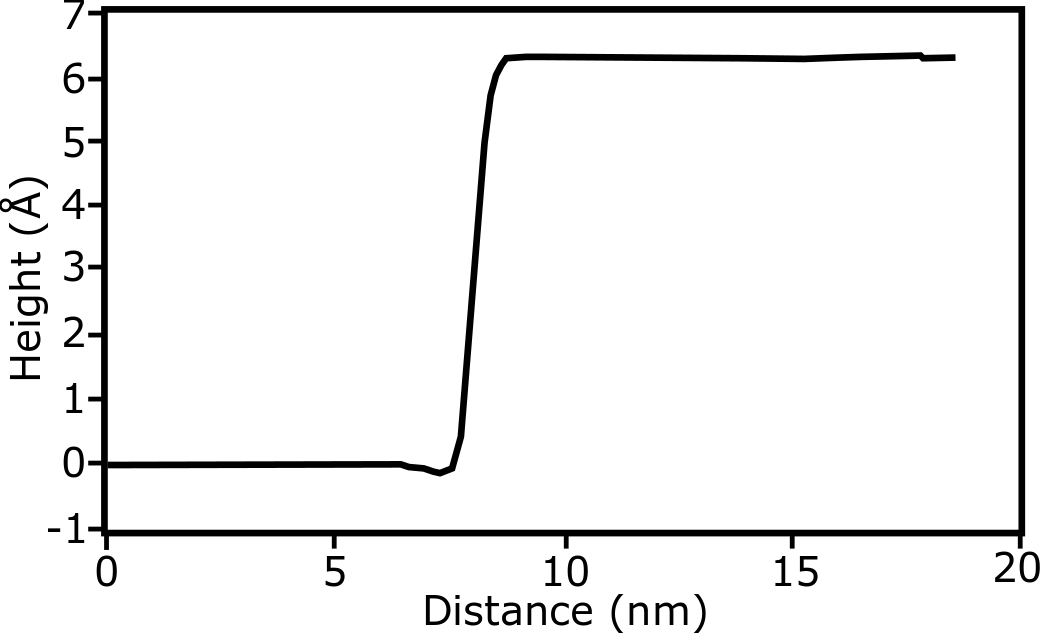}

\end{figure}
\noindent {\bf Fig. S1.} Left: STM image of the monolayer edge, with no evidence of solitons occurring in the bulk crystal. Right: Line cut showing a height difference of 6.3\AA\, across the edge, consistent with a single layer terrace in MoSe$_2$.
\\

\begin{figure}[h!]
\centering
\includegraphics[width=0.4\textwidth]{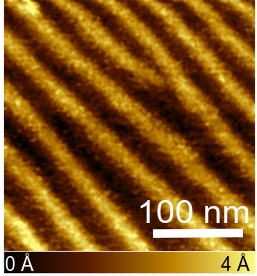}
\end{figure}
\noindent {\bf Fig. S2.} STM image of a dislocation in a striped incommensurate state. 
\\

\clearpage

\begin{figure}[h!]
\centering
\includegraphics[width=0.6\textwidth]{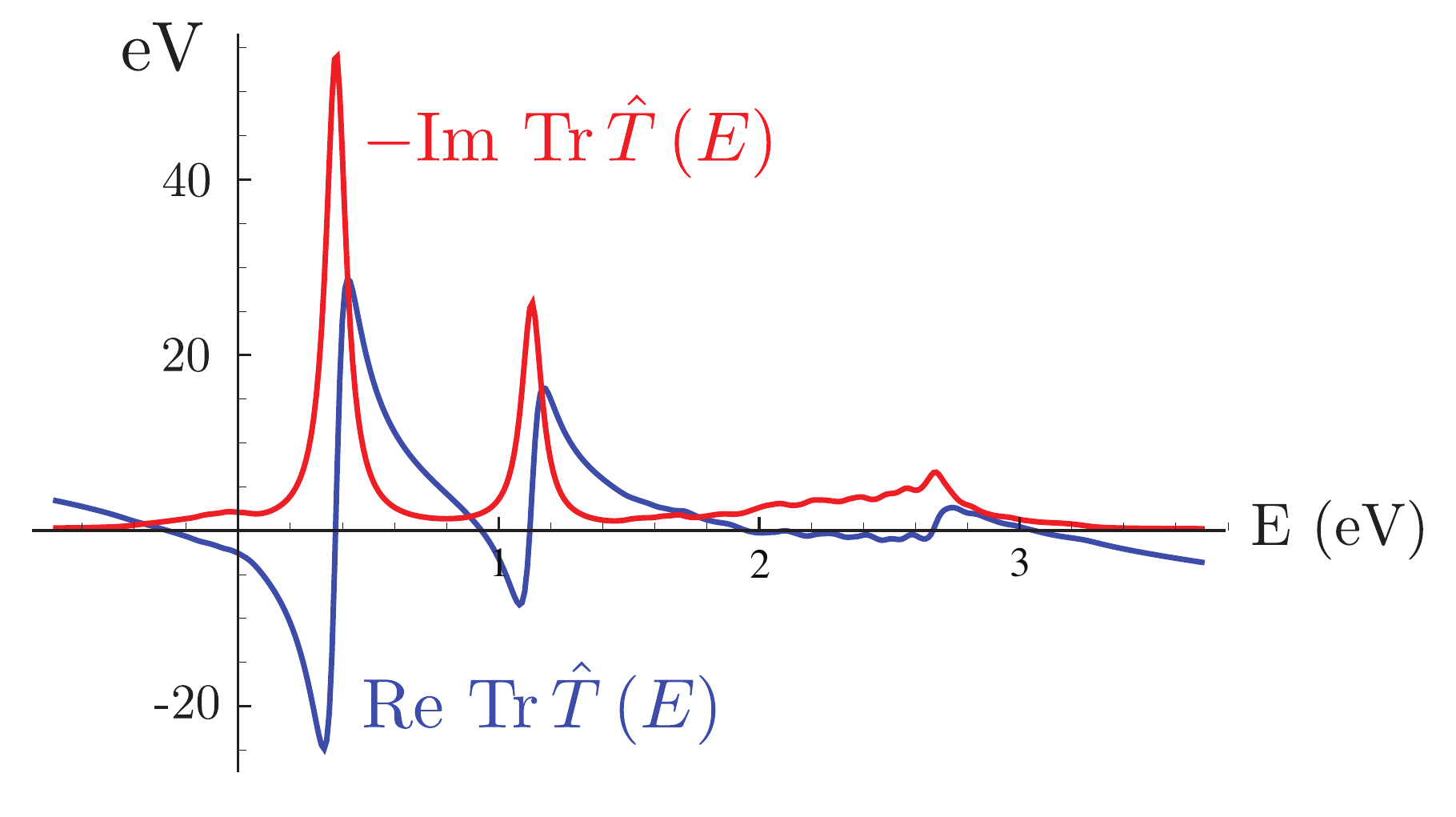}
\end{figure}
\noindent {\bf Fig. S3.} Resonances of the $T$-matrix in the \textit{strong scatterer} limit. We used the tight-binding parameters in \cite{31}, diagonalized $\hat{H}_{it}$ in $\mathbf{k}$-space and approximated the sum in crystal momenta implicit in the definition of $\hat{G}_0^R(E,0,0)$ by a mesh of 625 points within the Brillouin zone. We took a spectral width of $\delta=47$ meV.

\end{document}